\documentclass[aps,prb,showpacs,12pt]{revtex4}
\usepackage{amsmath,amssymb}
\usepackage{graphicx,latexsym}
\def\beq{\nopagebreak \begin{equation}}
\def\eeq{\end{equation}}
\def\denr{{n({\bf r})}}
\def\br{{\bf r}}

\begin{document}
\title{Functional form of the generalized gradient approximation for exchange: The PBE$\alpha$ functional.}
\author{Georg K. H. Madsen}
\affiliation{Dept. of Chemistry, University of Aarhus, DK-8000 {\AA}rhus C, Denmark}
\email{georg@chem.au.dk}
\date{\today}
\begin{abstract}
A new functional form for the exchange enhancement in the generalized gradient approximation within density functional theory is given. The functional form satisfies the constraints used to construct the Perdew-Burke-Ernzerhof (PBE) functional but can be systematically varied using one parameter. This gives the possibility to estimate the reliability of a computational result or to fit the parameter for a certain problem. Compared to other semi-empirical functionals, the present has the advantage that only one physically transparent parameter is used and that the fitted functional will obey the same exact conditions as PBE functional. Furthermore the simple form of the exchange enhancement means that oscillating terms in the exchange potential are avoided.
\end{abstract}
\maketitle

\section{Introduction}
Density functional theory (DFT) has made it possible to calculate ground state properties of even very large systems efficiently and accurately. As the exact functional is unknown, practical applications of DFT require an approximate exchange correlation energy, $E_{xc}[n]$.\cite{kohnsham} Of these, the local spin density approximation (LDA)\cite{kohnsham,vwn,pw92} has been successful despite its simplicity. The generalized gradient approximations (GGAs) are an attempt to improve on the LDA. It is possible to construct a GGA free of empirical parameters, known as the Perdew Burke Ernzerhof (PBE)-GGA, by demanding that the GGA obeys certain fundamental constraints.\cite{PBE} Numerical tests have shown that the PBE-GGA gives total energy dependent properties in good agreement with experiment.\cite{testpkzb,testTPSS} Consequently, the PBE-GGA functional has been extremely influential, both for performing actual calculations and as a basis for functionals involving higher derivatives and exact exchange.\cite{Jacob}

As both the density and the gradient can only be constant in the homogeneous electron gas limit, there can be no unique GGA\cite{PBEexample} and the constraints of the PBE-GGA are not sufficient to uniquely define the functional. The PBE functional form was based on a numerical GGA\cite{pw86,ggahole} where a model of the exchange correlation hole was constructed to satisfy known exact hole constraints. The constraints were satisfied using a sharp real space cutoff and a damping function, which were choices of the authors and different choices would lead to different functionals.\cite{ggahole} This was recently demonstrated by Wu and Cohen (WC),\cite{WC} who used the gradient expansion for slowly varying densities\cite{SvB} to construct a GGA with a functional form corresponding to a diffuse cutoff.\cite{WC} The WC-GGA damps the gradient enhancement, which results in improved equilibrium volumes for densely packed solids, but poorer exchange energies of atoms compared to the PBE.\cite{WC} Another modification of the PBE, the RPBE\cite{RPBE}, chooses a functional form with a larger gradient enhancement, which generally leads to very good atomic exchange energies but poor equilibrium volumes of solids.\cite{testpkzb} The PBE, RPBE and WC functionals differ only in the functional form of the exchange energy enhancement and all satisfy the same conditions as the PBE. 

A different approach to has been to construct GGAs as parameterized fits (See Kurth et al.\cite{testpkzb} for an overview). These semiempirical functionals are often precise for a certain set of compounds, but the parameters are rarely physically transparent and it is consequently not clear when they could fail. Furthermore they generally violate the known exact conditions satisfied by the PBE-GGA. This paper introduces a new functional form that brings the WC, PBE and RPBE exchange functionals onto a common ground and allows semiempirical functionals, that do not violate the constraints of the PBE-GGA, to be constructed. The shape of the functional is controlled by one physically transparent parameter and in the construction of the functional it is ensured that the exact conditions of PBE are not violated.

\section{Background}
\begin{figure*}
\includegraphics[width=.49\textwidth]{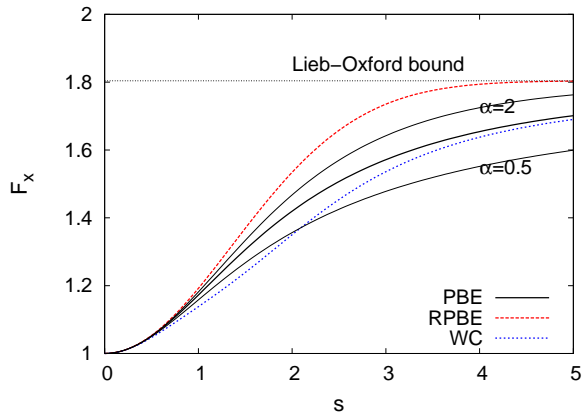}
\caption{(color online) Exchange enhancements, $F_x$, due to density gradients. The PBE, RPBE and WC functionals are shown with thick lines. Two PBE$\alpha$ functionals are shown with thin lines and marked with the $\alpha$ used.}
\label{fig:funcform}
\end{figure*}

The GGAs write the exchange energy density pr. particle as
\begin{equation}
\epsilon_{x}(n,s)=\epsilon_x(n) F_{x}(s)
\end{equation}
where $\epsilon_x(n)$ is the LDA exchange energy density pr. particle. $F_{x}$ is the enhancement factor due to density gradients and is dependent on the reduced density gradient, $s=\frac{|\nabla n(\br)|}{2(3\pi^2)^{1/3}\denr^{4/3}}$. The PBE enhancement factor is given as
\begin{equation}
F^{PBE}_x(x)=1+\kappa\biggl( 1-\frac{1}{1+x/\kappa}\biggr)
\label{eq:FxPBE}
\end{equation}
where $x=\mu p$ and $p=s^2$. The parameters of Eq.~(\ref{eq:FxPBE}), $\mu=0.2195$ and $\kappa=0.804$, are determined to ensure that the exchange gradient correction cancels that for PBE correlation as $s\rightarrow0$ and to ensure that the local Lieb-Oxford bound is obeyed.\cite{PBE} The RPBE functional differs in the functional form  
\beq
F^{RPBE}_x(x)=1+\kappa(1-e^{-x/\kappa}) 
\eeq
while the WC functional retains the functional form of PBE but introduces a complicated $x^{WC}=10/81 p+(\mu-10/81)pe^{-p}+\ln(1+cp^2)$.\cite{WC} The enhancement factors of WC, PBE and RPBE are shown in Fig.~\ref{fig:funcform}.

Obviously the WC, PBE and PRBE functionals satisfy the local Lieb-Oxford bound. By Taylor expanding $F_x$ for $p\rightarrow 0$ it is clear that all three exchange functionals have the correct first order expansion, $F_x=1+\mu p+O\left(p^2\right)$, and thus obey the cancellation condition:
\begin{gather}
F_x^{PBE} = 1+\mu p-\frac{\mu^2}{\kappa} p^2+\frac{\mu^3 }{\kappa^2}p^3+O\left(p^4\right) \\
F_x^{RPBE} = 1+\mu p-\frac{1}{2}\frac{\mu^2 }{\kappa}p^2+\frac{\mu^3 }{6 \kappa^2}p^3+O\left(p^4\right) \\
\begin{split}
F_x^{WC} = 1&+\mu p-\left(\frac{\mu^2}{\kappa}+\mu-b-c\right) p^2\\
            &+\left(\frac{\mu^3}{\kappa^2}+\frac{2 \mu^2-2 b \mu-2 c \mu}{\kappa}+\frac{\mu-b}{2}\right) p^3+O\left(p^4\right) \\
\end{split}
\end{gather}

The exchange enhancement, which will be named PBE$\alpha$, that will be presented here has the form
\begin{equation}
F_x^{PBE\alpha}=1+\kappa \left(1-\frac{1}{\left(1+x_1/(\kappa \alpha)\right)^\alpha}\right)
\end{equation}
where $\alpha$ is a parameter. The Taylor expansion 
\begin{equation}
F_x^{PBE\alpha}=1+\mu p-\frac{\alpha+1}{2 \alpha}\frac{\mu^2}{\kappa}p^2+\frac{\mu^3 (\alpha+2) (\alpha+1) p^3}{6 \kappa^2 \alpha^2}+O\left(p^4\right)
\label{eq:PBEaTaylor}
\end{equation}
shows that PBE$\alpha$ satisfies the cancellation condition for all $\alpha$. The PBE and RPBE are then special cases of PBE$\alpha=1$ and PBE$\alpha\!\rightarrow\!\infty$ respectively. The PBE$\alpha$ functional is shown for two different $\alpha$ in Fig.~\ref{fig:funcform} where it is seen how the size of the gradient enhancement can be controlled through the $\alpha$ parameter. Other attempts have been made at systematically varying the PBE functional form either by making an expansion of the enhancement coefficients\cite{mPBE} or by varying the $\kappa$ parameter.\cite{revPBE,kappavar} However, these functionals either violate the local Lieb-Oxford bound or can only ensure it through a fitting constraint.

\section{Discussion}
In the following, the influence of the parameter $\alpha$ will be discussed. All calculation have been performed using the WIEN2k code.\cite{WIEN2k} The basis sets and $k$-meshes have been converged and equations of states (EOSs) have been found by fitting the stabilized jellium EOS\cite{sjeos} to the calculated total energies. 

\subsection{Influence of $\alpha$}
The influence of the parameter $\alpha$ can be illustrated by comparing the body centered cubic (BCC) phase of Fe with the hexagonal close packed (HCP). The example is well studied and it is known that LDA wrongly predicts the non-magnetic FCC phase to be stable at zero pressure.\cite{Fe_GGA} EOS parameters have been calculated for BCC and HCP iron using LDA and PBE$\alpha$ for three choices of $\alpha$,  Table~\ref{tab:FeEOS}. Using the EOS parameters, BCC-HCP phase transition pressures of 14.0, 16.4 and 18.3 GPa are predicted for PBE$\alpha=0.8$, $\alpha=1.0$ and $\alpha=1.2$ respectively, in reasonable agreement with a recent experimental study which fixed the transition pressure between 14 and 16 GPa.\cite{Fe_Xray} 

The iron example shows how the $\alpha$ parameter can be interpreted. A small $\alpha$ favors the close packed structure (in the extreme case of the LDA the HCP structure is predicted to be stable at ambient pressure) while a large $\alpha$ favors an open structure. This also illustrates the dangers of the most obvious use of the PBE$\alpha$ functional: Fitting $\alpha$ for a certain problem. If one uses data set of densely packed structures and a small $\alpha$, one risks biasing the results towards the densely packed structures. If one uses a data set of open structures (or in the extreme: atoms) and obtain a large $\alpha$ one risks biasing the results towards the open structures. Often it is probably better to use the well tested PBE$\alpha=1.0$ functional. However, varying $\alpha$ will provide information on how sensitive the results obtained are too variations of the functional. Such a philosophy was also recently proposed in a statistical study of the exchange enhancement.\cite{BayDFT} The PBE$\alpha$ functional would probably better suited for this approach than the form actually used,\cite{BayDFT} which was similar to the mPBE.\cite{mPBE}

\begin{table}
\begin{ruledtabular}
\begin{tabular}{l l|c c c c}
     &                 & $E_0-E_0^{BCC}$ & $V_0$          & $B_0$ & $B_0'$ \\
     &                 &  (mRy/atom)     & (a$_0^3$/atom) & (GPa) & \\
\hline
BCC  & LDA              &   0.00          & 70.19          & 250.69 & 4.97 \\
     & PBE$\alpha=0.8$  &   0.00          & 76.18          & 190.69 & 5.29 \\
     & PBE$\alpha=1.0$  &   0.00          & 76.45          & 188.11 & 5.32 \\
     & PBE$\alpha=1.2$  &   0.00          & 77.03          & 182.34 & 5.33 \\
\hline
HCP  & LDA             &  -9.76          &  64.80         & 337.72 & 4.30 \\
     & PBE$\alpha=0.8$ &   6.15          &  68.81         & 288.06 & 4.52 \\
     & PBE$\alpha=1.0$ &   7.25          &  69.10         & 285.63 & 4.53 \\
     & PBE$\alpha=1.2$ &   8.10          &  69.30         & 282.70 & 4.54 \\
\end{tabular}
\end{ruledtabular}
\caption{Equation of state parameters for BCC and HCP iron.}
\label{tab:FeEOS}
\end{table}

\subsection{Limits of $\alpha$}
Another question would be whether there is an upper and an lower bound for $\alpha$. The good performance of the RPBE for atoms and WC for densely packed solids can give the impression that these functionals form the upper and lower bound of the gradient enhancement. The RPBE performs extremely well for the exchange energy of atoms which would indicate that there is no upper limit for $\alpha$. This is shown in Table~\ref{tab:noble_ex} where MARE decreases with $\alpha$.

\begin{table}
\begin{ruledtabular}
\begin{tabular}{l|c |c c c c c c c}
   &  X-OEP    & WC        &$\alpha=0.52$ &  $\alpha=1$ & $\alpha=2$ & $\alpha=5$  & $\alpha=20$  & RPBE   \\
\hline
He &   -1.0258 & -0.9805   & -0.9916      &    -1.0051  &   -1.0145  &   -1.0212   &   -1.0249  & -1.0262 \\
Ne &  -12.1050 & -11.8676  & -11.9597     &   -12.0275  &  -12.0716  &  -12.1015   &  -12.1176 &-12.1231 \\
Ar &  -30.1747 & -29.6846  & -29.8646     &   -29.9814  &  -30.0551  &  -30.1039   &  -30.1299 &-30.1388 \\
Kr &  -93.8330 & -92.8559  & -93.1862     &   -93.3769  &  -93.4925  &  -93.5671   &  -93.6060 &-93.6193 \\
Xe & -179.0635 & -177.5052 & -177.9762    &  -178.2368  & -178.3923  & -178.4914   & -178.5428 &-178.5602  \\
\hline
MARE &         &  1.98     &  1.37        & 0.85        & 0.50       &  0.26       & 0.18      & 0.16
\end{tabular}
\end{ruledtabular}
\caption{Atomic exchange energies in Ha. The X-OEP marks the optimized effective potential values taken from Kurth et al.\cite{testpkzb} MARE is the mean absolute relative error in \%.}
\label{tab:noble_ex}
\end{table}

If exponentially decaying densities in atoms constitute one extreme, the other extreme is the slowly varying density. For the slowly varying density the expansion of $F_x$ is known\cite{SvB} 
\begin{equation}
F_x^{SvB}=F_x=1+\frac{10}{81}p+\frac{146}{2025}q^2-\frac{73}{405}qp+Dp^2+O(\nabla^6)
\label{eq:SvB}
\end{equation}
where $D\approx0$ and $q=\frac{\nabla^2 n}{4k_F^2n}=\frac{\nabla^2 n}{4(3\pi^2)^{2/3}n^{5/3}}$. The expansion in Eq.~(\ref{eq:SvB}) was the inspiration for the WC functional,\cite{WC} where the Laplacian terms were avoided by introducing the approximation: $q\approx 2/3 p$. Introducing this approximation into Eq.~(\ref{eq:SvB}), one obtains $F_x^{SvB}\approx 1+10/81 p-0.0881207 p^2$. From Eq.~(\ref{eq:PBEaTaylor}) it is seen that if one sets $\alpha=0.52$ in the PBE$\alpha$ functional the same second order term is obtained. In Tables~\ref{tab:vol} and \ref{tab:bulk} the equilibrium volumes and bulk moduli of 18 solids\cite{WC,TPSS,testTPSS} calculated with the $PBE\alpha=0.52$ functional are given and compared to the results of the LDA, PBE, RPBE and WC functionals. It can be seen that the PBE$\alpha=0.52$ functional gives results that are similar to the WC functional, being slightly worse on the equilibrium volumes and slightly better on the bulk moduli. One cannot exactly reproduce the WC functional with the PBE$\alpha$ because of its different $x_2$ but both the WC and the PBE$\alpha=0.52$ functional were constructed to give a good description of the known gradient expansion of $E_x$ for a slowly varying density.\cite{SvB} The performance of PBE$\alpha=0.52$ thus confirms that this expansion is important for densely packed solids.\cite{WC}

\begin{table}
\begin{ruledtabular}
\begin{tabular}{l|c |c c c c c c c c}
Solid & $V_0^{exp}$ & $V_0^{LDA}$ & $V_0^{PBE}$ & $V_0^{RPBE}$ & $V_0^{WC}$ & $V_0^{PBE\alpha=0.52}$ \\
\hline
Li    & 141.83      & 128.32      & 136.68      & 142.10       & 138.37   & 134.13 \\
Na    & 255.48      & 224.17      & 249.50      & 266.49       & 249.48   & 241.37 \\ 
K     & 481.31      & 432.57      & 501.21      & 545.80       & 497.04   & 479.00 \\
Al    & 110.59      & 106.66      & 111.37      & 113.50       & 109.66   & 110.16 \\
C     &  76.57      &  74.50      &  77.03      &  78.13       &  75.89   &  76.36 \\
Si    & 270.11      & 266.22      & 274.41      & 281.58       & 270.79   & 273.59 \\
SiC   & 139.64      & 136.93      & 141.88      & 144.06       & 139.56   & 140.58 \\
Ge    & 304.61      & 300.20      & 322.97      & 332.33       & 309.20   & 316.80 \\
GaAs  & 303.96      & 297.43      & 320.53      & 330.42       & 306.63   & 314.20 \\
NaCl  & 295.48      & 275.22      & 312.64      & 337.12       & 300.86   & 300.10 \\
NaF   & 165.18      & 153.52      & 175.11      & 188.37       & 169.36   & 168.36 \\
LiCl  & 224.58      & 207.45      & 231.84      & 247.88       & 220.74   & 223.53 \\
LiF   & 108.79      & 100.70      & 113.23      & 120.31       & 108.72   & 109.41 \\
MgO   & 125.62      & 121.69      & 130.35      & 134.72       & 126.71   & 127.93 \\
Cu    &  78.91      &  73.56      &  80.58      &  83.44       &  76.74   &  78.80 \\
Rh    &  92.43      &  89.95      &  95.29      &  96.94       &  92.45   &  94.11 \\
Pd    &  98.62      &  95.82      & 103.19      & 105.93       &  99.05   & 101.43 \\
Ag    & 113.66      & 107.88      & 119.31      & 124.57       & 112.49   & 116.19 \\
\hline
MRE   &             &  -5.26      &   2.88      &   7.32       &  -0.03   & 0.52   \\
STD   &             &   3.30      &   2.76      &   4.22       &   1.70   & 2.56   \\
MARE  &             &   5.26      &   3.54      &   7.32       &   1.32   & 1.93   \\
\end{tabular}
\end{ruledtabular}
\caption{Volume of the primitive unit cell (in a$_0^3$). MRE is mean relative error. STD is the standard deviation of the relative error. MARE as in Table~\ref{tab:noble_ex}. All errors are in \%}
\label{tab:vol}
\end{table}

\begin{table}
\begin{ruledtabular}
\begin{tabular}{l|c |c c c c c c c c}
Solid & $B_0^{exp}$ & $B_0^{LDA}$ & $B_0^{PBE}$ & $B_0^{RPBE}$ & $B_0^{WC}$ & $B_0^{PBE\alpha=0.52}$ \\
\hline
Li    &   13.0      &  15.06      &  14.01      &  13.51       &  13.39     &  14.40 \\
Na    &    7.5      &   9.19      &   7.71      &   6.97       &   7.48     &   8.15 \\
K     &    3.7      &   4.49      &   3.62      &   3.35       &   3.42     &   3.82 \\
Al    &   79.4      &  82.68      &  76.49      &  74.43       &  79.08     &  77.90 \\
C     &   443       & 467.35      & 432.44      & 417.49       & 449.20     & 441.77 \\
Si    &   99.2      &  95.13      &  87.78      &  83.95       &  92.74     &  90.07 \\
SiC   &   225       & 228.70      & 211.76      & 204.53       & 220.72     & 216.42 \\
Ge    &   75.8      &  72.40      &  59.42      &  54.07       &  67.62     &  62.89 \\
GaAs  &   75.6      &  74.50      &  60.93      &  55.44       &  69.42     &  64.71 \\
NaCl  &   26.6      &  32.76      &  24.32      &  20.56       &  24.88     &  26.77 \\
NaF   &   51.4      &  63.65      &  46.73      &  40.14       &  47.08     &  51.12 \\
LiCl  &   35.4      &  42.04      &  31.81      &  26.22       &  35.46     &  35.20 \\
LiF   &   69.8      &  90.66      &  71.73      &  65.07       &  74.12     &  76.50 \\
MgO   &   165       & 173.41      & 147.76      & 135.70       & 156.54     & 154.80 \\
Cu    &   142       & 193.22      & 143.93      & 125.90       & 170.78     & 156.13 \\
Rh    &   269       & 320.43      & 263.74      & 245.63       & 295.16     & 276.47 \\
Pd    &   195       & 241.40      & 180.36      & 159.76       & 214.94     & 194.81 \\
Ag    &   109       & 146.51      &  96.64      &  76.98       & 126.06     & 109.79 \\
\hline
MRE   &             &  15.27      & -6.17       & -14.84       & 0.61  & -0.43 \\
STD   &             &  13.04      & 7.61        &   9.52       & 8.82  &  7.81 \\
MARE  &             &  16.39      & 7.81        &  15.27       & 6.80  &  5.57 \\
\end{tabular}
\end{ruledtabular}
\caption{Bulk moduli $B_0$ (in GPa). MRE, STD and MARE as in Table~\ref{tab:vol}.}
\label{tab:bulk}
\end{table}


\subsection{Potential}
\begin{figure*}
\includegraphics[width=.49\textwidth]{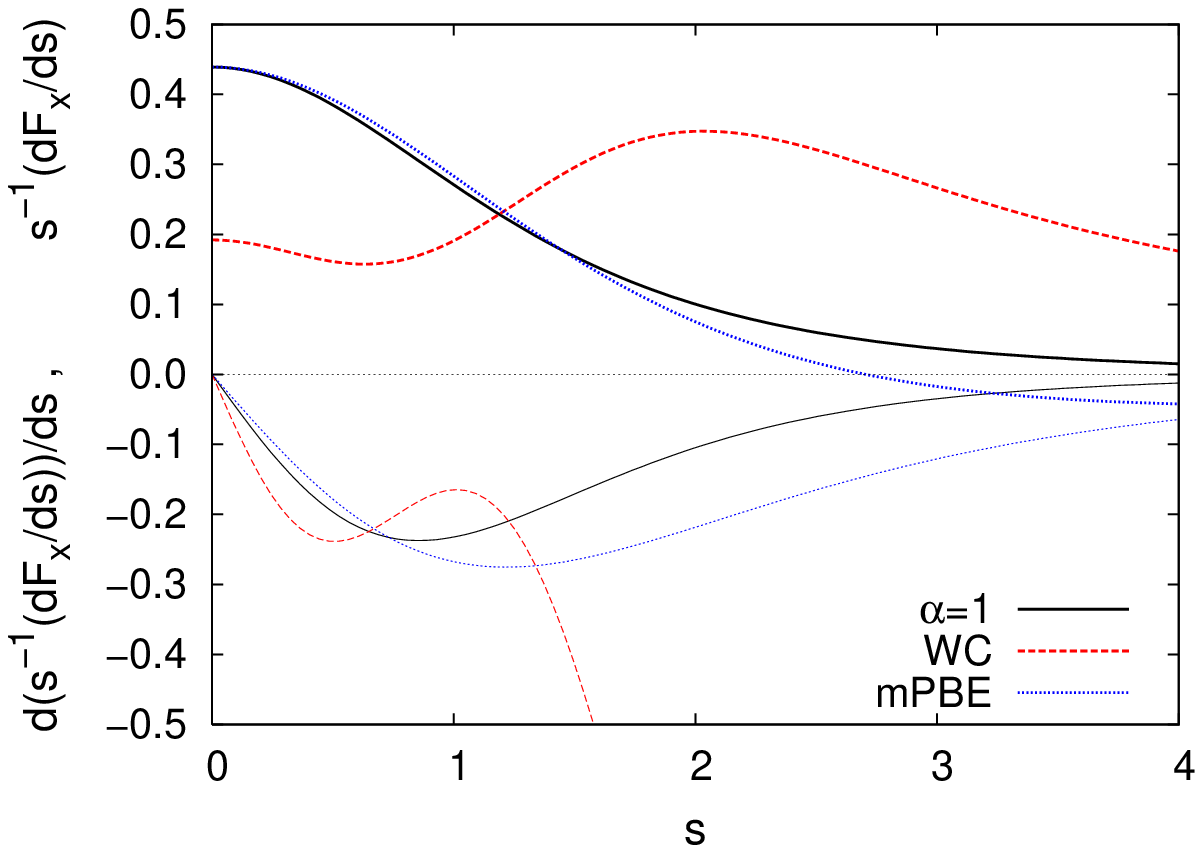}
\caption{(color online) Derivatives of the exchange enhancement that enter the exchange potential (Eq.~\ref{eq:xcpotential1}). Lines for PBE$\alpha=1$, WC and mPBE $F_x$ are shown. The thick lines are for the $s^{-1} \frac{dF_x}{ds}$ term and the thin lines for the $\frac{d}{ds}\biggr(s^{-1}\frac{dF_x}{ds}\biggl)$ term.}
\label{fig:partder}
\end{figure*}

Though often ignored in the construction of a functional, the resulting exchange-correlation potential is an important quantity in judging the quality of a functional.\cite{Burkeviral} According to Perdew and Wang,\cite{pw86} the exchange potential of a GGA can be calculated as
\beq
v_x=\frac{d E_x}{dn}=\epsilon_x \Biggr[\frac{4}{3} F_x - q\bigr(s^{-1} \frac{dF_x}{ds}\bigl)-(u-\frac{4}{3}s^3) \frac{d}{ds}\biggr(s^{-1}\frac{dF_x}{ds}\biggl) \Biggl]
\label{eq:xcpotential1}
\eeq
where $u=(2 k_F)^{-3}n^{-2} \nabla n \cdot \nabla|\nabla n|$. The potential thus depends on both the first and second derivative of the enhancement form. One advantage of the PBE$\alpha$ form compared to earlier attempts to vary the exchange enhancement is that its simpler form leads a smoother exchange potential. This is shown in Fig.~\ref{fig:partder} where the terms $s^{-1}\frac{dF_x}{ds}$ and $\frac{d}{ds}\biggr(s^{-1}\frac{dF_x}{ds}\biggl)$ from the PBE$\alpha=1$, the WC and the mPBE $F_x$ are shown. It can be seen that the WC terms, due to the complicated form of $x_2$, have an oscillating behavior while the mPBE converges very slowly to zero. The slow convergence of the mPBE is particularly unpleasant as this term is multiplied by the $4/3s^3$, which diverges for large $s$. Though GGAs in general do not have the correct $1/r$ asymptotic decay, the mPBE does not correct this problem, and the effects is an artifact of a poorly constructed functional form.

\section{Conclusion}
A new functional form for GGAs has been proposed, which can be systematically
varied by the parameter $\alpha$. One could thereby obtain a functional that
combines the virtues of semi-empirical GGAs, high precision for a specific set
of compounds, with the virtues of PBE, obeying the same exact constraints as
the PBE. Compared to other semi-empirical functionals it is an
advantage that only one parameter is used and that the $\alpha$ has clear
interpretation, which can be used to check
whether results are biased towards densely packed or open systems. One further
advantage of the PBE$\alpha$ is that the simple construction of the functional
form leads to a smooth exchange potential. 

It is clear, that the limited amount of information in the gradient of the density makes it impossible to make a GGA functional that works equally well for all cases. On higher rungs of the DFT ladder,\cite{Jacob} such as the meta-GGAs, this could be possible. One way to include the extra information is make the $x$ dependent on more variables, as in the recently proposed TPSS meta-GGA.\cite{TPSS} A different route could be to make $\alpha$ be determined by the density and not be the same every where in space. One could e.g. use that for an exponentially decreasing density $q/p=1$ to identify regions where $\alpha$ should be large. An other idea which does not involve higher derivatives was put forward by Wu and Cohen,\cite{WC} who suggested making the exchange enhancement dependent directly on the density ($F_x(n,s)$).

A subroutine implementing the PBE$\alpha$ can be found at http://www.chem.au.dk/$\sim$webuorg/georg.


\begin{thebibliography}{10}

\bibitem{kohnsham}
W. Kohn and L.~J. Sham, Phys. Rev. {\bf 140},  A1133  (1965).

\bibitem{vwn}
S.~J. Vosko, L. Wilk, and M. Nusair, Can. J. Phys. {\bf 58},  1200  (1980).

\bibitem{pw92}
J.~P. Perdew and Y. Wang, Phys. Rev. B {\bf 45},  13244  (1992).

\bibitem{PBE}
J.~P. Perdew, K. Burke, and M. Ernzerhof, Phys. Rev. Lett. {\bf 77},  3865
  (1996).

\bibitem{testpkzb}
S. Kurth, J.~P. Perdew, and P. Blaha, Int. J. Quantum Chem. {\bf 75},  889
  (1999).

\bibitem{testTPSS}
V.~N. Staroverov, G.~E. Scuseria, J. Tao, and J.~P. Perdew, Phys. Rev. B {\bf
  69},  075102  (2004).

\bibitem{Jacob}
J.~P. Perdew, A. Ruzsinszky, J. Tao, V.~N. Staroverov, G.~E. Scuseria, and
  G.~I. Csonka, J. Chem. Phys. {\bf 123},  62201  (2005).

\bibitem{PBEexample}
J.~P. Perdew, M. Ernzerhof, A. Zupan, and K. Burke, J. Chem. Phys. {\bf 108},
  1522  (1998).

\bibitem{pw86}
J.~P. Perdew and Y. Wang, Phys. Rev. B {\bf 33},  R8800  (1986).

\bibitem{ggahole}
J.~P. Perdew, K. Burke, and Y. Wang, Phys. Rev. B {\bf 54},  16533  (1996).

\bibitem{WC}
Z. Wu and R.~E. Cohen, Phys. Rev. B {\bf 73},  235116  (2006).

\bibitem{SvB}
P.~S. Svendsen and U. von Barth, Phys. Rev. B {\bf 54},  17402  (1996).

\bibitem{RPBE}
B. Hammer, L.~B. Hansen, and J.~K. N\o{}rskov, Phys. Rev. B {\bf 59},  7413
  (1999).

\bibitem{mPBE}
C. Adamo and V. Barone, J. Chem. Phys. {\bf 116},  5933  (2002).

\bibitem{revPBE}
Y. Zhang and W. Yang, Phys. Rev. Lett. {\bf 80},  890  (1998).

\bibitem{kappavar}
E.~L. Peltzer~y Blanc{\'a}, C.~O. Rodr{\'i}guez, J. Shitu, and D.~L. Novikov,
  J. Phys.: Condens. Matter {\bf 13},  9463  (2001).

\bibitem{WIEN2k}
P. Blaha, K. Schwarz, G.~K.~H. Madsen, D. Kvasnicka, and J. Luitz, {\em WIEN2k,
  An Augmented Plane Wave Plus Local Orbitals Program for Calculating Crystal
  Properties. ISBN 3-9501031-1-2}, Vienna University of Technology, Austria,
  2001.

\bibitem{sjeos}
A.~B. Alchagirov, J.~P. Perdew, J.~C. Boettger, R.~C. Albers, and C. Fiolhais,
  Phys. Rev. B {\bf 63},  224115  (2001).

\bibitem{Fe_GGA}
L. Stixrude, R.~E. Cohen, and D.~J. Singh, Phys. Rev. B {\bf 50},  6442
  (1994).

\bibitem{Fe_Xray}
F. Baudelet, S. Pascarelli, O. Mathon, J.~P. Iti{\'e}, A. Polian, M. d'~Astuto,
  and J.~C. Chervin, J. Phys.: Condens. Matter {\bf 17},  957  (2005).

\bibitem{BayDFT}
J.~J. Mortensen, K. Kaasbjerg, S.~L. Frederiksen, J.~K. N{\o}rskov, J.~P.
  Sethna, and K.~W. Jacobsen, Phys. Rev. Lett. {\bf 95},  216401  (2005).

\bibitem{TPSS}
J. Tao, J.~P. Perdew, V.~N. Staroverov, and G.~E. Scuseria, Phys. Rev. Lett.
  {\bf 91},  146401  (2003).

\bibitem{Burkeviral}
K. Burke, F.~G. Cruz, and K.-C. Lam, J. Chem. Phys. {\bf 109},  8161  (1998).

\end{thebibliography}

\end{document}